\documentclass[twocolumn,twoside,slac_two]{revtex4}
\usepackage{graphicx}
\usepackage{fancyhdr}
\usepackage{amssymb}
\begin{document}

\title{The CTA Observatory}

\author{R. M. Wagner}
\affiliation{Max-Planck-Institut f\"ur Physik, D-80805 Munich, Germany} 
\affiliation{Excellence Cluster ``Universe'', D-85748 Garching, Germany}
\author{E. Lindfors}
\affiliation{Tuorla Observatory, Department of Physics and Astronomy, University of Turku, FIN-21500 Piikki\"o, Finland}
\author{S. Wagner}
\affiliation{Landessternwarte Heidelberg, D-69117 Heidelberg, Germany}
\author{A. Sillanp\"a\"a}
\affiliation{Tuorla Observatory, Department of Physics and Astronomy, University of Turku, FIN-21500 Piikki\"o, Finland}
\author{for the CTA Consortium}

\begin{abstract}
In recent years, ground-based very-high-energy (VHE; $E \gtrapprox 100$~GeV)
$\gamma$-ray astronomy has experienced a major
breakthrough with the impressive astrophysical results obtained mainly by the
current generation experiments like H.E.S.S., MAGIC, MILAGRO and
VERITAS. 
The ground-based Imaging
Air Cherenkov Technique for detecting VHE $\gamma$-rays has matured, and a fast
assembly of inexpensive and robust telescopes is possible. The goal for the
next generation of instruments is to increase their sensitivity by a factor
$\gtrsim 10$ compared to current facilities, to extend the accessible
$\gamma$-ray energies from a few tens of GeV to a hundred TeV, and to improve
on other parameters like the energy and angular resolution (improve the point-spread function by a factor $4-5$ w.r.t. current instruments). 
The Cherenkov Telescope Array (CTA) project is an initiative to build the next generation ground-based
$\gamma$-ray instrument, will serve as an observatory to a wide astrophysics
community. 
I discuss the key
physics goals and resulting design considerations for CTA,
the envisaged technical solutions chosen, and
the organizational and operational
requirements for operating such a large-scale facility as well as the specific
needs of VHE $\gamma$-ray astronomy. 
\end{abstract}

\maketitle

\thispagestyle{fancy}


\section{INTRODUCTION}\label{sec:intro}
Very-high energy (VHE) $\gamma$-rays are produced in nonthermal processes in
the universe, namely in galactic objects like pulsars, pulsar-wind nebulae,
supernova remnants (SNR), binary systems containing compact objects, or OB
associations. Among the extragalactic VHE $\gamma$-ray sources are active
galactic nuclei (AGN), particularly blazars and radio-galaxies, and starburst
galaxies. Galaxy clusters and gamma-ray bursts are also potential, although not
yet discovered, sources of VHE $\gamma$ rays.  Apart from the astrophysics of
specific astronomical objects, $\gamma$-ray astronomy can be used to search for
the annihilation of dark matter particles, and for studying the transparency
and history of the universe. Further fundamental physics searches, like for
the violation of Lorentz invariance, can be performed.  For recent reviews, see, e.g.,
\cite{Aharonian:2008zz,Buckley:2008}.

Upon reaching the Earth's atmosphere, VHE $\gamma$-rays interact with
atmospheric nuclei and generate electromagnetic showers. The showers extend
over several kilometers in length and few tens to hundreds of meters in width, and their
maximum is located at $8-12$~km altitude, in case of vertical
incidence.
At VHE, the shower particles are stopped high up in the atmosphere, and can not be directly detected at ground.  A sizeable
fraction of the charged secondary shower particles, mostly electrons and
positrons in the shower core, move with ultra-relativistic speed and emit
Cherenkov light.  This radiation is mainly concentrated in the near UV and optical band and
therefore passes mostly unattenuated to the ground, with minor losses due to Rayleigh
and Mie scattering and Ozone absorption. Imaging atmospheric Cherenkov telescopes reflect the Cherenkov
light onto multi-pixel cameras that record the shower images. 
The technique was pioneered by the Whipple
experiment, which first detected the Crab Nebula in TeV $\gamma$-rays, in 1989.

\section{CURRENT FACILITIES}
Currently, the world largest ground-based IACTs are H.E.S.S., MAGIC and
VERITAS.

  H.E.S.S. is an array of 4 identical 12-m diameter telescopes,
located in Namibia and operating since 2003. A fifth telescope of 28~m
diameter, is under construction in the center of the array, and its
completion is foreseen for 2010. 

MAGIC,  has been operated since 2004 as
17-m diameter single-dish telescope on the Canary Island if La Palma, Spain.
Despite the use of a single reflector does not guarantee the sensitivity of an array, its large 
reflector surface of 239 m$^2$ allowed
to reach  the lowest energy threshold among the ground-based instruments,
enabling for the first time observation below
100~GeV with this imaging air Cherenkov technique.
A second MAGIC telescope is now fully operational, and the
use of the stereoscopic observation technique will allow MAGIC
to reach a significantly improved sensitivity. 

A more recent
experiment was started in the Arizona desert in USA, following the
successful experience of the Whipple experiment. VERITAS has soon reached
the expected performance, with a sensitivity comparable to H.E.S.S., and is
starting to collect important scientific results.

\section{TOWARDS A PRECISION GAMMA-RAY ASTRONOMY}
Despite the achievements of current-generation Cherenkov
telescopes \cite{Aharonian:2008zz}, there are limitations that
future instruments will need to overcome: current instruments are sensitive in
an energy range of $\gtrsim 80$~GeV$-$50~TeV. At the low energy end,
systematic limitations come from the background from atmospheric hadronic (and electronic)
showers. At the high end, the limit is statisticsi due to the too small
collection areas at the high (multi-10 TeV) energies). 
telescopes is limited to a typical field of view (FOV) of $3-5^\circ$~diameter,
as is the angular resolution, currently around a few arcmin. Also, current
facilities are rather poorly automatized.
From a physics point of view, there are strong arguments to improve in the
following aspects: decrease the energy threshold to few tens of GeV; acquire
sensitivity beyond 50~TeV; increase sensitivity in the core range
(100~GeV$-$50~TeV); improve energy and angular resolution. Cherenkov Telescope
Array (CTA) is a next-generation ground-based project which aims at
implementing all these improvements.

\section{CHERENKOV TELESCOPE ARRAY}

\begin{figure}
\centering
\includegraphics[width=0.85\linewidth]{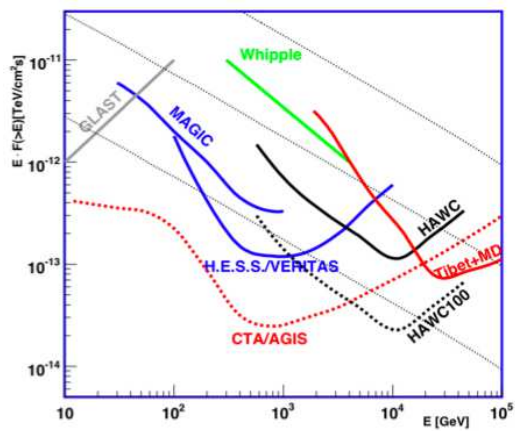}
\caption{\label{fig:cta_sens}Integral sensitivity for a Crab-like
  spectrum for several current imaging air Cherenkov telescope facilities and expected for CTA and 
the Advanced Gamma-ray Imaging System; cf. \cite{Buckley:2008}), a similar US-based project
  (5$\sigma$, 50~h)
  and Fermi-LAT (5$\sigma$, 1~yr).}
\end{figure}

The success of ground based $\gamma$-ray astronomy experiments in recent years has
brought nearly all scientists working in the field in Europe together to design
and promote CTA. This instrument will achieve superior
sensitivity by deploying a large number of Cherenkov telescopes of different
sizes covering a large area on the ground for high detection rates. CTA
foresees improvement of sensitivity of factor 5-10 in the current energy domain
(somewhat below 100 GeV to some 10 TeV) and will extend the energy range from
10 GeV to about 100 TeV (Fig. \ref{fig:cta_sens}). The observatory will consist of two arrays: a southern
hemisphere array, which allows deep investigation of galactic sources and of
the central part of our Galaxy, but also for the observation of extragalactic
objects. The northern hemisphere array is dedicated mainly to northern
extragalactic objects. Obviously the arrays will not be only restricted to pure
astrophysical observations, but will also make contributions to the field of
particle physics and cosmology.

\begin{figure*}
\centering
\includegraphics[width=\linewidth]{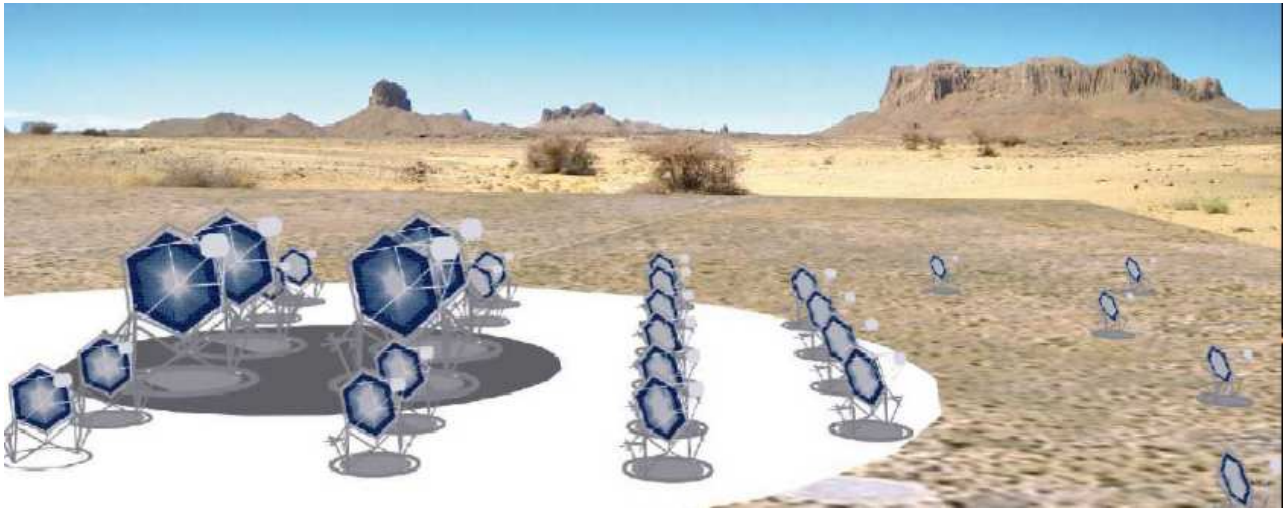}
\caption{\label{fig:cta_artist}Artist's view of the composite CTA observatory.
The area covered amounts to $1-10$~km$^2$.}
\end{figure*}

CTA will be operated as an open, proposal-driven facility analogous to optical
observatories, that shall be available for all scientists from those countries
that contribute to the construction and operation of the observatory. It is
foreseen to follow the practice of other major, successfully operating
observatories (e.g. the European Southern Observatory) and announce calls for
proposals on regular intervals which will be peer-reviewed by a changing group
of international experts. Based on experience of current experiments and other
ground-based observatories, different classes of proposals (targeted, surveys,
time-critical, Target of Opportunity and regular programs) are foreseen. User
support will be provided via a data centre, in the form of standard processing
of data and access to the standard MC simulations and analysis pipelines used
in data processing.

\section{SCIENTIFIC MOTIVATION}
\subsection{Low-energy physics (sub-50 GeV)}
MAGIC has opened the field of ground-based sub-100~GeV $\gamma$-ray astronomy
\cite{Aliu:2008}. Observations with higher sensitivity in this region will have
important consequences for galactic and extragalactic physics.  By studying the
sub-50 GeV energy band, CTA could provide the final answer to the acceleration
mechanism in pulsars.  A high sensitivity in the low energy regime is also
vital for studying AGNs, which typically exhibit rather steep power-law spectra
(due to the increasing suppression of $\gamma$-ray flux with energy by the
extragalactic background light). Further, low-energy sensitivity is vital to
complement the catalog of \emph{Fermi}-LAT detected $\gamma$-ray emitters with
a possibly larger significance than obtainable by LAT at its high-energy end.
This will provide a unique way to understand the nature of the tens of yet
unidentified \emph{Fermi}-LAT detected sources~\cite{Abdo:2009mg}. 
Also, $\gamma$-rays in the $E<100$~GeV regime will help the broad-band modelling and thus
 to distinguish between the two basic classes of acceleration models, i.e.,
leptonic and hadronic acceleration models.

\subsection{High-energy physics (above 50~TeV)}
Typical astrophysical $\gamma$-ray spectra have bimodal distributions with one
peak at lower energies due to synchrotron emission, and a second peak at
higher energies due to inverse Compton scattering of VHE electrons on
seed infrared/optical photons. For galactic objects, one may expect to
observe VHE power-law $\gamma$-ray spectra with cutoffs due to intrinsic
mechanisms. $\gamma$-rays from distant blazars suffer a severe
attenuation after pair production with local IR-UV photons of the
extragalactic background light (EBL). In all current IACT data, the
evidences for spectral cutoffs and steepening 
are rather poor. There is no simple justification for this.
CTA will explore this region with unprecedented
sensitivity. This will allow to understand the
acceleration mechanism in galactic objects like SNRs, and
discriminating hadronic/leptonic models. The high energy region is also important for identifying 
PeVatrons, i.e. cosmic-ray accelerators to TeV/PeV energies.
Detecting outbursts (``flares'') from distant AGNs at
super-TeV energies would allow to increase the prospects for firm limits on
Lorentz invariance violation.

\subsection{Core energy region (0.1$-$50~TeV)}
CTA will provide an increase of the sensitivity in this core energy region by
at least a factor 10 as compared to the current IACTs, reaching a level of
$10^{-3}$~C.U. 
sensitivity in this range. 
(The differential photon flux of the Crab nebula between 60 GeV and 9 TeV is \cite{magiccrab} 
${\rm d}F/{\rm d}E=(6.0\pm0.2)10^{-10} (E/300\,{\rm GeV})^{a+b\log_{10}(E/300\, {\rm GeV})} \,\mathrm{TeV}^{-1} \mathrm{cm}^{-2} \mathrm{s}^{-1}$ with $a=-2.31\pm0.06$, $b=-0.26\pm0.07$.)
This will promote $\gamma$-ray astrophysics
to a \emph{$\gamma$-ray astronomy}. In fact, for the first
time, CTA will allow for a full VHE sky survey, with approx. a thousand new
VHE~$\gamma$-ray sources expected to be detected. 

An increase in sensitivity will, by reducing the required observation times, 
allow more follow-up observations and higher
time resolution of variable sources.
The current telescopes are sensitive
enough to detect variations on timescalew of minutes. CTA will allow 
a sub-minute resolution, and thus to
understand the complex phenomena of $\gamma$-ray flares, directly
connected to the acceleration mechanisms and the local
environment. 
Morphological studies will profit from reduced required observation
times. These are of importance to study spatially extended $\gamma$-ray emitters, like SNRs.
An increase of the angular resolution by a factor $4-5$, down to 0.02 arcmins (current theoretical limits are discussed in \cite{angres}), will reduce source confusion and improve collaboration with instruments observing
at other wavelengths.

\section{GENERAL DESIGN IDEAS}
maintain an overall high technical performance, the CTA concept (Fig. \ref{fig:cta_artist}) is based on few
general ideas: Increase the array from currently 4 (H.E.S.S., VERITAS) to
$\approx 100$ telescopes; distribute them over a large area ($1-10$~km$^2$);
make use of telescopes of $2-3$ different sizes; take advantage of  well-proven
technology of current IACTs; high automatization and remote operation; run
array as observatory and open the facility to the astronomy and astrophysics
community.

\section{General technical ideas on CTA}

An increase in sensitivity over the full energy range can only
be achieved by combining many telescopes distributed over a large area of at
least 1~km$^2$ and using telescopes of $2-3$ different sizes: several medium
size telescopes (MST) of 12~m, few large size telescopes (LST) of 24~m
diameter, and probably several small size telescopes (SST) of 7~m diameter. The
number of the telescopes, their size, configuration and the overall
performance are under investigation.

The LSTs detect sub-100 GeV photons thanks to their large reflective area.
Technologically, they are the most challenging telescope type. A design is
currently under development.
Several tens of MSTs will perform the $\gamma$-ray detection in the core energy
region. Those telescopes will be based on the experience gained with the
H.E.S.S. and MAGIC telescopes. The main goal is to reduce cost and maintenance
efforts. The MSTs constitute the core of the array, and will also perform the
fundamental task of vetoing the LST triggers to reduce hadronic background. The
MST design is currently studied and the construction of a prototype is expected
in few years from now.
In case 3 different telescope sizes are required, several tens of SSTs will
complete the array to perform the super-TeV search, increasing the effective
collection area of the array. Very simple in construction, contributing only a
small percentage to the cost of the full array.  

\begin{figure}
\centering
\includegraphics[width=.7\linewidth]{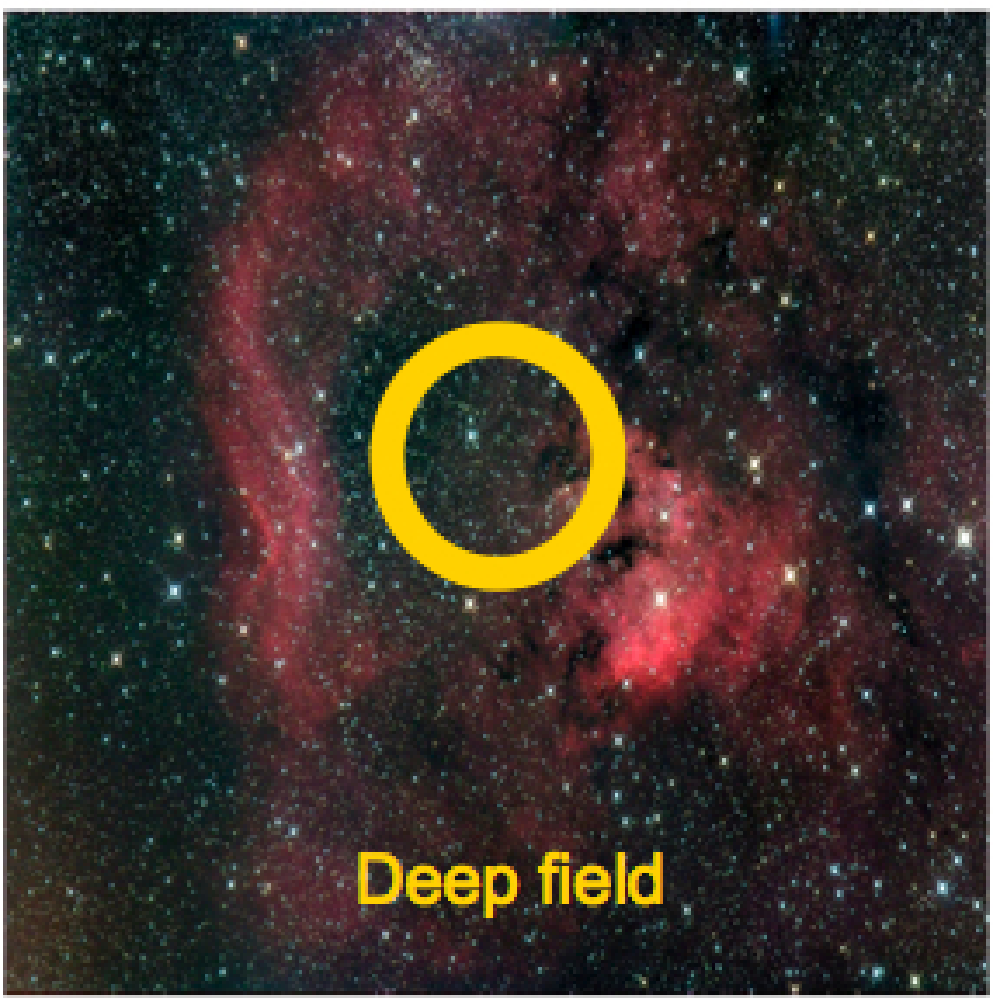}\\
\includegraphics[width=.7\linewidth]{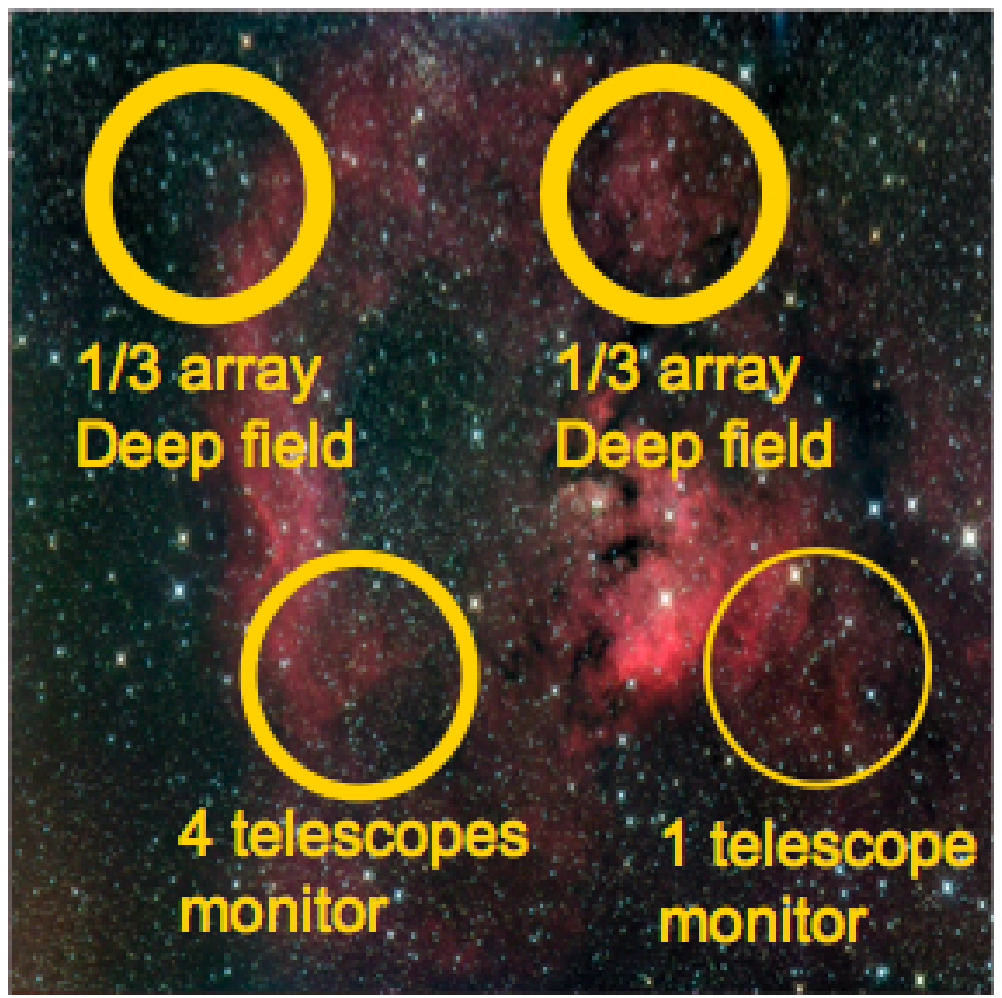}\\
\includegraphics[width=.7\linewidth]{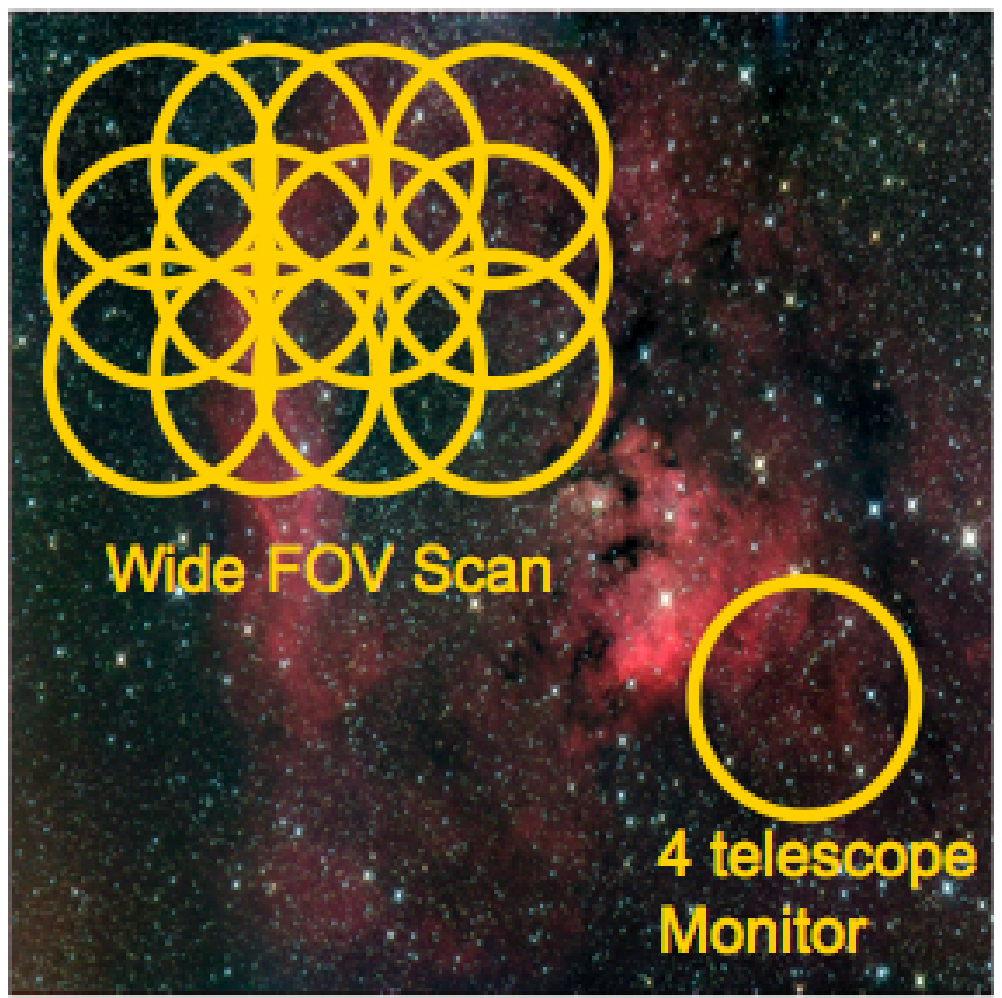}\\
\caption{\label{fig:cta_modes}Possible operation modes for CTA. From
  left to right: deep field, flexible and wide field modes. More
  details in the text.}
\end{figure}

The trigger systems will support different operation modes (Fig. \ref{fig:cta_modes}).  In the ``deep
field'' mode, all telescopes will be pointed to the same sky position to
maximize the sensitivity. In a more flexible mode, parts of the telescopes
could point to different positions, with few telescopes making follow-up
observation of single sources as, e.g., to monitor blazar activity. Finally,
the array can be operated in a ``wide-FOV'' mode, to perform an all-sky scan in
a time-efficient way at a moderate sensitivity.

\section{The CTA consortium}
CTA is a partnership between the H.E.S.S. and MAGIC collaborations and several
European institutes, with recent interests from more institutions world-wide.
Activities are coordinated with AGIS (Advanced Gamma-ray Imaging System; cf.
\cite{Buckley:2008}), a similar US-based project.  The consortium comprises $>
70$ institutes from 17 countries, involving more than 400 scientists.

For the current design phase, CTA is organized in several work packages:
Management, Physics, Monte Carlo, Site, Mirror, Telescope, Focal-Plane
Instrumentation, Electronics, Atmospheric Transmission and Calibration,
Observatory, Data, and Quality Assurance.
The telescope design and component prototyping are expected to be completed in
2011 and 2012, respectively.  Array prototypes could exist by 2012/13, and the
construction of the full array and partial operation could be started in
2014/15.  

\section{SPECIFIC NEEDS OF VHE GAMMA-RAY ASTRONOMY}

Like the major optical and radio observatories the CTA arrays will be situated
in remote locations, which means that the operation of the array should be as
robotic as possible. On the other hand, the detection technique of Cherenkov
telescopes require operation of high voltages, which makes the robotic
operations challenging. It is also foreseen that the observations of a single
source are distributed over several nights and on same night several sources
from different proposals will be observed. Therefore the onsite operations i.e.
the observations and maintainance will be handled by the onsite staff rather
than visiting astronomers.

Another challenge will be the data rate from the foreseen ~100 telescopes. The remote location will most likely not allow the real-time transfer of the raw data, but the data will rather be pre-processed onsite. 
The layout of the CTA observatory will consist of operations centre/centres,
data centre, managment and user community.

\section*{ACKNOWLEDGMENTS}
We want to thank all our colleagues from the CTA consortium for the tremendous work being done during the design study. The support of the involved national funding agencies and of the European community is gratefully acknowledged, as is the support by the H.E.S.S. and MAGIC collaborations and the interested parties from the US and Japan.
R.M.W.'s research is support in part by the DFG Cluster of Excellence
``Origin and Structure of the Universe''.

\end{document}